\documentclass[twocolumn,preprintnumbers,amsmath,amssymb]{revtex4}
\usepackage{latexsym,amssymb,amsthm,amsmath,epsfig}
\usepackage[title]{appendix}
\usepackage{hyperref}
\hypersetup{
    colorlinks=true,
    linkcolor=blue,
    filecolor=blue,      
    urlcolor=blue,
}

\begin{document}

\title{Dynamics and multiqubit entanglement in distant resonators}
\author{Muhammad Waseem$^{1}$}
\email{mwaseem328@gmail.com}
\author{Muhammad Nehal Khan$^{2}$}
\author{Shahid Qamar$^{2,3}$}
\affiliation{$^{1}$ Karachi Institute of Power Engineering College, Pakistan Institute of Engineering and Applied Sciences (PIEAS), Karachi 75780, Pakistan}
\affiliation{$^{2}$ Department of Physics and Applied Mathematics, PIEAS, Nilore, Islamabad 45650, Pakistan}
\affiliation{$^{3}$ Center for Mathematical Sciences, PIEAS, Nilore, Islamabad 45650, Pakistan}

\date{\today }

\begin{abstract}
We consider the dynamics of the photon states in distant resonators coupled to a common bus resonator at different positions.
The frequencies of distant resonators from a common bus resonator are equally detuned.
These frequency detunings are kept larger than the coupling strengths of each resonator to the common bus resonator to satisfy the dispersive interaction regime.
In the dispersive regime, we show that the time dynamics of the system evolve to an arbitrary W-type state in a single step at various interaction times.
Our results show that a one-step generation of arbitrary W-type states can be achieved with high fidelity in a system of superconducting resonators.

keywords: Time dynamics, W-state, Entanglement, distant Resonators, Dispersive interaction, photon catch, photon release.
\end{abstract}

\maketitle

\section{introduction}
The coherent interaction of matter with quantized fields within high reflective resonators or Cavity known as cavity Quantum Electrodynamics (QED) provides a powerful platform for quantum information processing~\cite{Bennett,haroch,kimble} and quantum networks~\cite{duan}. 
Attentions have been focused on the dynamics of coupled cavities for encoding the quantum information~\cite{walther,ogden,gma,ycl,fazal,fazal22,fazal24,eli}.
For example, the dynamics of the coupled-cavity array system with each cavity containing single two-level atoms~\cite{fazal24} or ensembles of atoms~\cite{fazal} revealed evidence for the formation of W-state.

In an alternative scenario of circuit-QED~\cite{wallraff}, there are some interesting studies, for example, the realization of quantum logic gates~\cite{waseemc,ming50,ming54,waseemgate,cp2016,cp2018,tong,ling}, quantum algorithms~\cite{wgrover}, microwave pulse or photon storage~\cite{mohsin,qad,cpsun,switch}, and entanglements~\cite{tian,ming47,ming52,ming,ming2,ming49,ming53}.
In some recent studies, the generation of W-state entanglement by encoding the quantum information in artificial qubits coupled to a microwave transmission line resonator was proposed~\cite{helmer,gao,huang,wei,kang,daltan}. 
In a system of artificial qubits coupled to a superconducting resonator, three-qubit W-state was generated through a dispersive collective quantum non-demolition measurement~\cite{helmer} and with resonant interaction~\cite{gao,wei}.

In the present paper, we theoretically examine the time dynamics and realization of W-type states in distant quantum resonators coupled to a common bus resonator at different positions with dispersive interaction. In the dispersive regime, we consider equal frequency detuning of each distant resonator from a common bus resonator, which is larger than their coupling strengths to common bus resonator.
We show that the time dynamics of the photon states $\left\vert 0\right\rangle$ and $\left\vert 1 \right\rangle$ (which serves as a qubit) in distant resonators lead to an arbitrary $n$-qubit W-state~\cite{ba,ba2}
\begin{equation}
\left\vert W \right\rangle_{n}=C_{1} \left\vert 100...0 \right\rangle + C_{2} \left\vert 010..0 \right\rangle + ...,+ C_{n}\left\vert 000..1 \right\rangle
\label{wn}
\end{equation}
in a single step at various interaction times.
Here, $\left|C_{1}\right|^{2} = \left|C_{2}\right|^{2} = ... = \left|C_{n}\right|^{2} = 1/n$.
We also discuss the experimental feasibility of our scheme by considering the effect of the finite lifetime of the photon in the resonator, inhomogeneous coupling of distant resonators to common bus resonator, direct resonator-resonator coupling, and mixed initial state. 
Our results show that a one-step generation of arbitrary W-state can be achieved with high fidelity. 

The W-type entangled state presented here may have potential applications in quantum sensing networks~\cite{sens1,sens2} and optimal processing of quantum information~\cite{ba}.
Our scheme of generating a W-type entangled state is general and can be applied to a different kind of coupled quantum system that follows the ladder algebra of quantum harmonic oscillators such as optical cavities and superconducting resonators.
Furthermore, encoding quantum information tasks in artificial atoms coupled to resonators is not required unlike some earlier studies~\cite{ming53,gao,wei}, which may make the realization simpler.

The paper is organized as follows. 
In Sec.II, we discuss the physical model and derive its effective Hamiltonian in the dispersive interaction regime.
In Sec.III, we study the time dynamics of the photon states population in distant resonators and show the evidence of W-state appearance.
In Sec.IV, we discuss the experimental feasibility in superconducting resonators and examine the fidelity of W-state against the experimental imperfections. Finally, we conclude our results in Sec.V. 

\begin{figure}[tbp]
 \includegraphics[width=3.25 in]{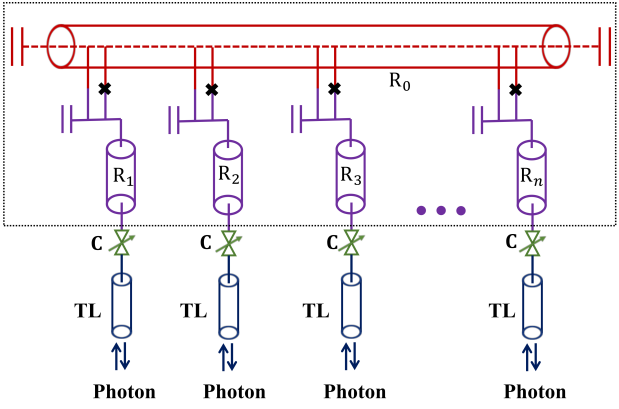}
\caption{Schematic diagram of the $n$ distant superconducting resonators $R_{j}$ ($j=1,2,3,...,n$) coupled to a common superconducting resonator $R_{0}$ through coupling strength $g_{j}$. The common superconducting resonator acts as a quantum bus. 
Here, superconducting resonator $R_{j}$ can catch and release a microwave photon from or to the superconducting Transmission Line (TL) by turning on the coupling capacitance $C$. 
The subsystem shown in the dotted box is the setup to study the time dynamics and for generating W-states on microwave photons in the different distant superconducting resonators.}
\label{fig:sys}       
\end{figure}

\section{Physical model}
Consider a one dimensional common resonator $R_{0}$ of frequency $\Omega_{0}$ with Hamiltonian $H_{R_{0}}=\hbar \Omega_{0} a^{\dag} a$. Here, $a^{\dag}$ and $a$ are the creation and annihilation operators of the common resonator $R_{0}$ holding the commutator property $[a,a^{\dag}]=1$. 
Next, we consider one dimensional distant resonators $R_{j}$ ($j=1,2,3,...$) coupled to a common resonator $R_{0}$ at different positions having Hamiltonian $H_{R_{j}}=\hbar \omega_{j} b_{j}^{\dag} b_{j}$. 
Here, $\omega_{j}$ is the frequency of $j^{th}$ distant resonator and $b_{j}^{\dag}$ ($b_{j}$) is the creation (annihilation) operators, which follows the property of $[b_{j^{\prime}},b_{j}^{\dag}]=\delta_{j j^{\prime}}$. 
It is assumed that the spacing between each distant resonator $R_{j}$ is large enough to avoid interaction among them.
Our proposed physical system using superconducting resonators is shown in Fig~\ref{fig:sys}.
The coupling strength $g_{j}$ between the $j^{th}$ resonator $R_{j}$ to common resonator $R_{0}$ can be tuned by external flux through the Superconducting Ring Couplers~\cite{ming45,ming57}.
The catch and release of microwave photon states $\left\vert 0\right\rangle_{j}$ and $\left\vert 1 \right\rangle_{j}$ from Transmission Line (TL) into superconducting resonator $R_{j}$ can be performed by turning on the capacitance $C$~\cite{ming45}.
The dotted line box in Fig.~\ref{fig:sys} shows the subsystem composed of superconducting resonators $R_{j}$ and $R_{0}$, which is used to generate the entanglement on the microwave photon states.

We first consider four-resonator subsystem composed of $R_{j}$ ($j=1,2,3$) and $R_{0}$ in dispersive regime. 
In the interaction picture, the Hamiltonian of four-resonator subsystem under rotating wave approximation can be written as (here, we assume $\hbar=1$)~\cite{ming57}
\begin{equation}
H=\sum_{j=1}^{3} g_{j} \left(a^{\dag} b_{j} e^{-i \Delta_{j} t} + a b^{\dag}_{j} e^{i \Delta_{j} t} \right),
\label{H}
\end{equation}
where, $g_{j}$ is the coupling strength of resonator $R_{j}$ to common resonator $R_{0}$ at different positions.
The detuning of the each resonator $R_{j}$ from common resonator $R_{0}$ is given by $\Delta_{j}=\Omega_{0}-\omega_{j}$.

We consider the dispersive regime such that $g_{j}/\Delta_{j} \ll 1$. This regime can be studied using a second-order perturbation expansion in $g_{j}/\Delta_{j}$.
Under this condition, it is convenient to rewrite the Hamiltonian $H$ of the four-resonator subsystem in Schr\"{o}dinger's picture such that
\begin{equation}
H_{s}=H_{0,s}+H_{I,s},
\label{Hs}
\end{equation}
with
\begin{equation}
H_{0,s}=\Omega_{0} a^{\dag}a + \sum_{j=1}^{3} \omega_{j} b^{\dag}_{j} b_{j}
\label{nonint}
\end{equation}
and
\begin{equation}
H_{I,s}=\sum_{j=1}^{3} g_{j}\left(a^{\dag} b_{j} + a b^{\dag}_{j} \right).
\label{int}
\end{equation}
The interaction part of the Hamiltonian given by Eq.~(\ref{int}) describes the coupling of three resonators to a common bus resonator $R_{0}$.
More insight into the dispersive regime can be gained by adiabatically eliminating the degree of freedom of common resonator $R_{0}$, which results in effective interaction among the distant resonators $R_{j}$.
For this purpose, we apply the unitary transformation~\cite{wallraff}, $U H_{s} U^{\dag}$, with
\begin{equation}
U=\mathrm{exp} \left[\sum_{j=1}^{3} \frac{g_{j}}{\Delta_{j}} \left(a b^{\dag}_{j}-a^{\dag} b_{j} \right)\right].
\label{U1}
\end{equation}
Expanding the above transformation to the second order in the small parameter $g_{j}/\Delta_{j}$, we can get the dispersive Hamiltonian given by
\begin{equation}
H^{\prime} = H_{0}+H_{I},
\label{hp}
\end{equation}
where, 
\begin{subequations}
\label{hi}
\begin{align}
H_{0} & =\Omega_{0}^{\prime} a^{\dag}a+\sum_{j=1}^{3} \omega_{j}^{\prime} b^{\dag}_{j} b_{j},\\
H_{I} & =\chi_{12}(b_{1}b^{\dag}_{2}+b^{\dag}_{1} b_{2})+\chi_{13}(b_{1}b^{\dag}_{3}+b^{\dag}_{1} b_{3}) \nonumber\\
			& +\chi_{23}(b_{2}b^{\dag}_{3}+b^{\dag}_{2} b_{3}).
\end{align}
\end{subequations}
The frequencies of the resonators $R_{j}$ are Lamb-shifted to $\omega_{j}^{\prime}=\omega_{j}+g_{j}^{2}/\Delta_{j}$, where $j=1,2,3$.
Similarly, the frequency of resonator bus $R_{0}$ is Lambed-shifted to $\Omega_{0}^{\prime}=\Omega_{0}- \sum_{j=1}^{3} g_{j}^{2}/\Delta_{j}$. For completeness, we give detailed derivation of Eq.~(\ref{hi}) in Appendix~\ref{A}.
The dispersive couplings between the resonator pairs of $R_{j}$ are given by
\begin{subequations}
\begin{align}
\chi_{12}& =\frac{g_{1} g_{2}}{2 \Delta_{1} \Delta_{2}} \left(\Delta_{1}+\Delta_{2}\right),\\
\chi_{13}& =\frac{g_{1} g_{3}}{2 \Delta_{1} \Delta_{3}} \left(\Delta_{1}+\Delta_{3}\right),\\
\chi_{23}& =\frac{g_{2} g_{3}}{2 \Delta_{2} \Delta_{3}} \left(\Delta_{2}+\Delta_{3}\right).
\end{align}
\label{couplings}
\end{subequations}
The Hamiltonian $H^{\prime}$ in the interaction picture, can be obtained using $H_{\rm eff} = e^{i H_{0} t} H_{I} e^{-i H_{0} t}$, and is given by
\begin{align}
H_{\rm eff} & = \chi_{12}b_{1}b^{\dag}_{2} e^{i \delta_{12} t}+\chi_{13} b_{1}b^{\dag}_{3} e^{i \delta_{13} t} \nonumber\\
						& + \chi_{23} b_{2}b^{\dag}_{3} e^{i \delta_{23} t} +H.c.,
\label{heff}
\end{align}
where, $\delta_{12}=\omega_{1}^{\prime}-\omega_{2}^{\prime}$, $\delta_{13}=\omega_{1}^{\prime}-\omega_{3}^{\prime}$ and $\delta_{23}=\omega_{2}^{\prime}-\omega_{3}^{\prime}$.
Equation~(\ref{heff}) describes the controllable interaction between the resonators $R_{1}$, $R_{2}$, and $R_{3}$.
In what follows, we study the time dynamics of the photon population in each resonators $R_{j}$ using the effective Hamiltonian given by~Eq.(\ref{heff}), which produces W-entangled state of photon states $\left\vert 0 \right\rangle$ and $\left\vert 1 \right\rangle$.
\begin{figure}[tbp]
 \includegraphics[width=3.5 in]{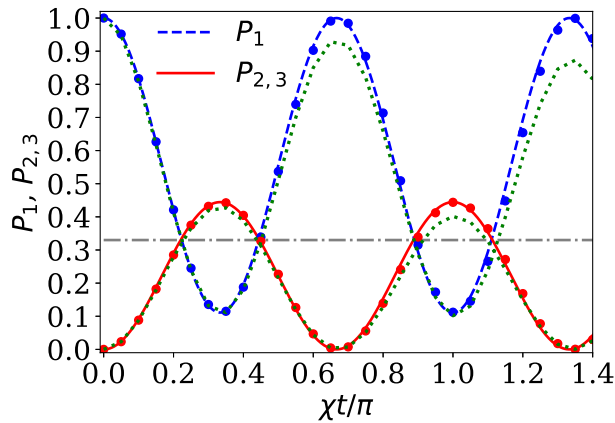}
\caption{Illustration for the time evolution of the photon population in four-resonator subsystem. The system is assumed to be initially in state $\left\vert 1_{1} 0_{2} 0_{3} \right\rangle \left\vert 0_{0} \right\rangle$ under the condition of identical coupling strengths $g$ and resonators having frequency $\omega_{r}$.
Here, dashed-blue curve represents the population of photons $P_{1}$ in resonator $R_{1}$ (for state $\left\vert 1_{1} 0_{2} 0_{3} \right\rangle \left\vert 0_{0} \right\rangle$), while solid red curve denotes the population of photons $P_{2,3}$ in resonator $R_{2}$ (for state $\left\vert 0_{1} 1_{2} 0_{3} \right\rangle \left\vert 0_{0} \right\rangle$) and resonator $R_{3}$ (for state $\left\vert 0_{1} 0_{2} 1_{3} \right\rangle \left\vert 0_{0} \right\rangle$). 
Filled circles indicate the resulting dynamics obtained from the ab initio Hamiltonian in Eq.~(\ref{Hs}).
Here, we choose parameters $\omega_{r}=2 \pi \times 5.75$~GHz, $\Omega_{0}=2 \pi \times 6.75$~GHz, $g=2 \pi \times 50$~MHz, and $\kappa=0$~MHz. Horizontal axis shows the operation time in unit of $\pi / \chi$.
The horizontal dashed-dotted line marks the point of interactions at which the dynamics of the system can nearly produce three-qubit W-state on photon states.
The dotted-green curve shows the damping of population oscillation in the presence of photon decay $\kappa=0.5$~MHz from the resonators.}
\label{fig:pop3}
\end{figure}

\section{Time dynamics and multiqubit entanglement}
In order to study the time dynamics of photon state population, we consider the subspace composed of photon states $\left\vert 1_{1} 0_{2} 0_{3} \right\rangle$, $\left\vert 0_{1} 1_{2} 0_{3} \right\rangle$ and $\left\vert 0_{1} 0_{2} 1_{3} \right\rangle$.
In this subspace, the state of the four-resonator subsystem is a linear combination of three states
\begin{equation}
\left\vert \psi (t) \right\rangle  = C_{1} (t) \left\vert \psi_{1} \right\rangle +  C_{2}(t) \left\vert \psi_{2} \right\rangle +  C_{3}(t) \left\vert \psi_{3} \right\rangle.
\label{psi}
\end{equation}
Here, $\left\vert \psi_{1} \right\rangle= \left\vert 1_{1} 0_{2} 0_{3} \right\rangle \left\vert 0_{0} \right\rangle $, 
$\left\vert \psi_{2} \right\rangle= \left\vert 0_{1} 1_{2} 0_{3} \right\rangle \left\vert 0_{0} \right\rangle$, 
and $\left\vert \psi_{3} \right\rangle= \left\vert 0_{1} 0_{2} 1_{3} \right\rangle \left\vert 0_{0} \right\rangle$.
The fourth qubit with subscript $0$ represents the vacuum photonic state of common resonator $R_{0}$.

The time evolution of the state $\left\vert \psi (t) \right\rangle$ of the four-resonator subsystem under effective Hamiltonian given by Eq.~(\ref{heff}) can be obtained by solving the Schr\"{o}dinger's Equation
\begin{equation}
\frac{\partial \left\vert \psi (t) \right\rangle }{\partial t} = -i H_{\rm eff} \left\vert \psi (t) \right\rangle
\label{swe}
\end{equation}
Using $\left\vert \psi (t) \right\rangle$ in Eq.~(\ref{swe}), the probability amplitudes can be expressed in the form of following coupled differential equations:
\begin{subequations}
\label{rate}
\begin{align}
\dot{C_{1}}& =-i\left[\chi_{12} e^{-i \delta_{12} t}C_{2}+\chi_{13} e^{i \delta_{13} t} C_{3}\right]\\
\dot{C_{2}}& =-i\left[\chi_{12} e^{i \delta_{12} t}C_{1}+\chi_{23} e^{-i \delta_{23} t} C_{3}\right]\\
\dot{C_{3}}& =-i\left[\chi_{13} e^{-i \delta_{13} t}C_{1}+\chi_{23} e^{i \delta_{23} t} C_{2}\right].
\end{align}
\end{subequations}
The initial state of four-resonator subsystem is assumed to be $\left\vert 1_{1} 0_{2} 0_{3} \right\rangle \left\vert 0_{0} \right\rangle$, which shows that there is a single photon in resonator $R_{1}$.
When resonators $R_{1}$, $R_{2}$ and $R_{3}$ are resonant to each other with $\omega_{r}=\omega_{1}=\omega_{2}=\omega_{3}$, than each resonator will be equally detuned from common resonator with detuning $\Delta=\Delta_{1}=\Delta_{2}=\Delta_{3}=\Omega_{0}-\omega_{r}$.
Assuming identical couplings i.e., $g=g_{1}=g_{2}=g_{3}$ to a common resonator, we have zero detuning values $\delta_{12}=\delta_{13}=\delta_{23}=0$ and identical dispersive couplings between the resonators pairs $R_{j}$ as $\chi_{12}=\chi_{13}=\chi_{23}=\chi$ (see Eq.~(\ref{couplings})).
Under this condition, the state of four-resonator subsystem evolves at time $t$ with coefficients 
\begin{subequations}
\label{amp}
\begin{align}
C_{1} (t)& =\frac{1}{3} \left[2 \cos \chi t + \cos 2 \chi t\right] \nonumber\\
				 & + \frac{i}{3} \left[2 \sin \chi t - \sin 2 \chi t \right],\\
C_{2} (t)& =C_{3}(t)=\frac{1}{3} \left[- \cos \chi t + \cos 2 \chi t\right] \nonumber\\
				 & - \frac{i}{3} \left[\sin \chi t + \sin 2 \chi t \right],
\end{align}
\end{subequations}
here, $\chi=g^{2}/\Delta$.
We first plot the population of photon states in resonator $R_{1}$, $R_{2}$ and $R_{3}$ as a function of operation time in dimensionless units $\chi t / \pi$ as shown in Fig.~\ref{fig:pop3}. The time evolution of photon state population can be obtained using~\cite{fazal24}
\begin{equation}
 P_{\rm m} = \left\langle \psi_{m} \left| \rho (t) \right| \psi_{\rm m} \right\rangle=C_{\rm m}C_{\rm m}^{*},
\label{pop}
\end{equation}
where $m=1,2,3$ and $\rho(t)=   \left\vert \psi (t) \right\rangle  \left\langle \psi (t) \right| $.
In Fig.~\ref{fig:pop3}, dashed-blue curve denotes the time evolution of the population $P_{1}$ of state $\left\vert 1_{1} 0_{2} 0_{3} \right\rangle \left\vert 0_{0} \right\rangle$ while solid-red curve denote the time evolution of the population $P_{2,3}$ of state $\left\vert 0_{1} 1_{2} 0_{3} \right\rangle \left\vert 0_{0} \right\rangle$ and $\left\vert 0_{1} 0_{2} 1_{3} \right\rangle \left\vert 0_{0} \right\rangle$. 
In order to get a comparison of the resulting dynamics obtained from the effective Hamiltonian~Eq.~(\ref{heff}), we also show the results of our numerical simulations obtained from the ab initio Hamiltonian in Eq.~(\ref{Hs}), which are shown by filled circles in Fig.~\ref{fig:pop3}.
Here, we choose parameters $\omega_{r}=2 \pi \times 5.75$~GHz, $\Omega_{0}=2 \pi \times 6.75$~GHz, and $\kappa = 0$~MHz. The coupling strength $g=2 \pi \times 50$~MHz~\cite{ming57} is chosen to satisfy the rotating wave approximation in Eq.~(\ref{H}).

The dynamics of the system shown in Fig.~\ref{fig:pop3} clearly reflect the transfer of photon population among the resonators $R_{1}$, $R_{2}$, and $R_{3}$.
The oscillation period of the photon populations depends on the dispersive coupling $\chi$. 
When there is a single photon in resonator $R_{1}$, Fig.~\ref{fig:pop3} shows that the dynamic of the system can evolve to a three-photon W-state at $\chi t / \pi \approx 0.22, 0.44, 0.88, 1.10$ with $\left|C_{1}\right|^{2} \approx \left|C_{2}\right|^{2} \approx \left|C_{3}\right|^{2} \approx 0.33$, just in a single step. 
During the entire time dynamics, the photon state population in resonator $R_{2}$ and $R_{3}$ oscillates with an equal magnitude due to the identical coupling strengths and frequencies. It is in-phase with each other while out of phase with the population in resonator $R_{1}$.

It may be mentioned here that for the case of coupled-cavity array system~\cite{fazal}, the probabilities of excitation (photon)-transfer was observed as high as 100$\%$. 
In distant resonator case considered in this study, the photon state population $P_{1}$ is never fully transferred ($P_{1} \neq 0$ throughout the dynamics) to the resonators $R_{2}$ and $R_{3}$ as can be seen in Fig.~\ref{fig:pop3}.
For interaction time other than $\chi t / \pi \approx 0.22, 0.44, 0.88, 1.10$, the condition $\left|C_{1}\right|^{2} \neq (\left|C_{2}\right|^{2} \approx \left|C_{3}\right|^{2})$ still holds. 
Therefore, it is clear that three-qubit entangled states with unequal probability amplitudes can still be generated except for time at which $P_{1}\rightarrow 1$ and $P_{2,3}\rightarrow 0$ as shown in Fig.~\ref{fig:pop3}. 
For example, choosing the operation time $\chi t=\pi / 3$, the final state evolves to
\begin{equation}
\left\vert \psi_{3} \right\rangle_{\pi / 3}  = \frac{1+i \sqrt{3}}{6} (\left\vert 1_{1} 0_{2} 0_{3} \right\rangle-2\left\vert 0_{1} 1_{2} 0_{3} \right\rangle - 2 \left\vert 0_{1} 0_{2} 1_{3} \right\rangle).
\label{w2}
\end{equation}
The subscript $\pi / 3$ indicates the operation time $\chi t$ at which time dynamics evolves to an entangled state. Similarly, for $\chi t=\pi$, the final state is given by
\begin{equation}
\left\vert \psi_{3} \right\rangle_{\pi}  = \frac{1}{3} (-\left\vert 1_{1} 0_{2} 0_{3} \right\rangle+2\left\vert 0_{1} 1_{2} 0_{3} \right\rangle + 2 \left\vert 0_{1} 0_{2} 1_{3} \right\rangle).
\label{w1}
\end{equation}
Both $\left\vert \psi_{3} \right\rangle_{\pi/3}$ and $\left\vert \psi_{3} \right\rangle_{\pi}$ are three-qubit entangled state of photons in three distant resonators $R_{1}$, $R_{2}$ and $R_{3}$.
Note that for both states $\left|C_{1}\right|^{2}=1/9$ and $\left|C_{2}\right|^{2}=\left|C_{3}\right|^{2}=4/9$. 
The behavior of the system symmetrically changes if a single photon is initially present in resonator $R_{2}$ i.e., $P_{2} \rightarrow P_{1}$ and $P_{1}=P_{3}$, as a result, W-state entanglement still preserved.
This situation is in contrast to coupled-cavity array system~\cite{fazal} where entanglement can disappear when initial excitation changes.

\begin{figure}[tbp]
\includegraphics[width=3.5 in]{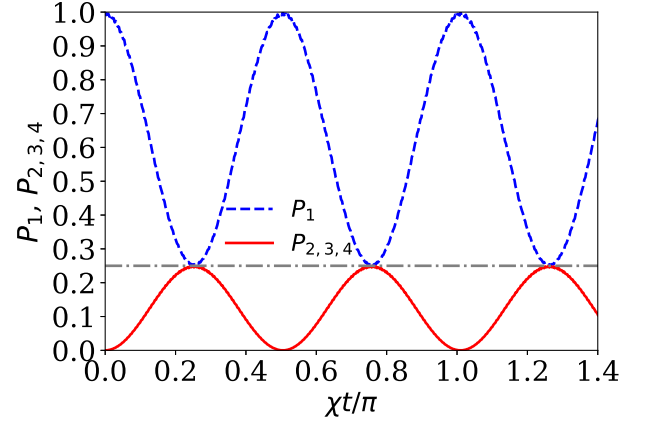}
\caption{Illustration for the time evolution of the photon population in five-resonator subsystem, which is assumed to be in the initial state $\left\vert 1_{1} 0_{2} 0_{3} 0_{4} \right\rangle \left\vert 0_{0} \right\rangle$.
Dashed-blue curve represents the population of photons $P_{1}$ in resonator $R_{1}$ while solid red curve denotes the population of photons $P_{2,3,4}$ in resonator $R_{2}$, $R_{3}$, and $R_{4}$.
The horizontal dashed-dotted line marks the point of interactions at which the dynamics of the system can nearly produce four-qubit W-state on photon states. The parameters adopted here are similar to Fig.~\ref{fig:pop3}.}
\label{fig:pop4}
\end{figure}
\begin{figure}[tbp]
\begin{tabular}{@{}cccc@{}}
\includegraphics[width=3.25 in]{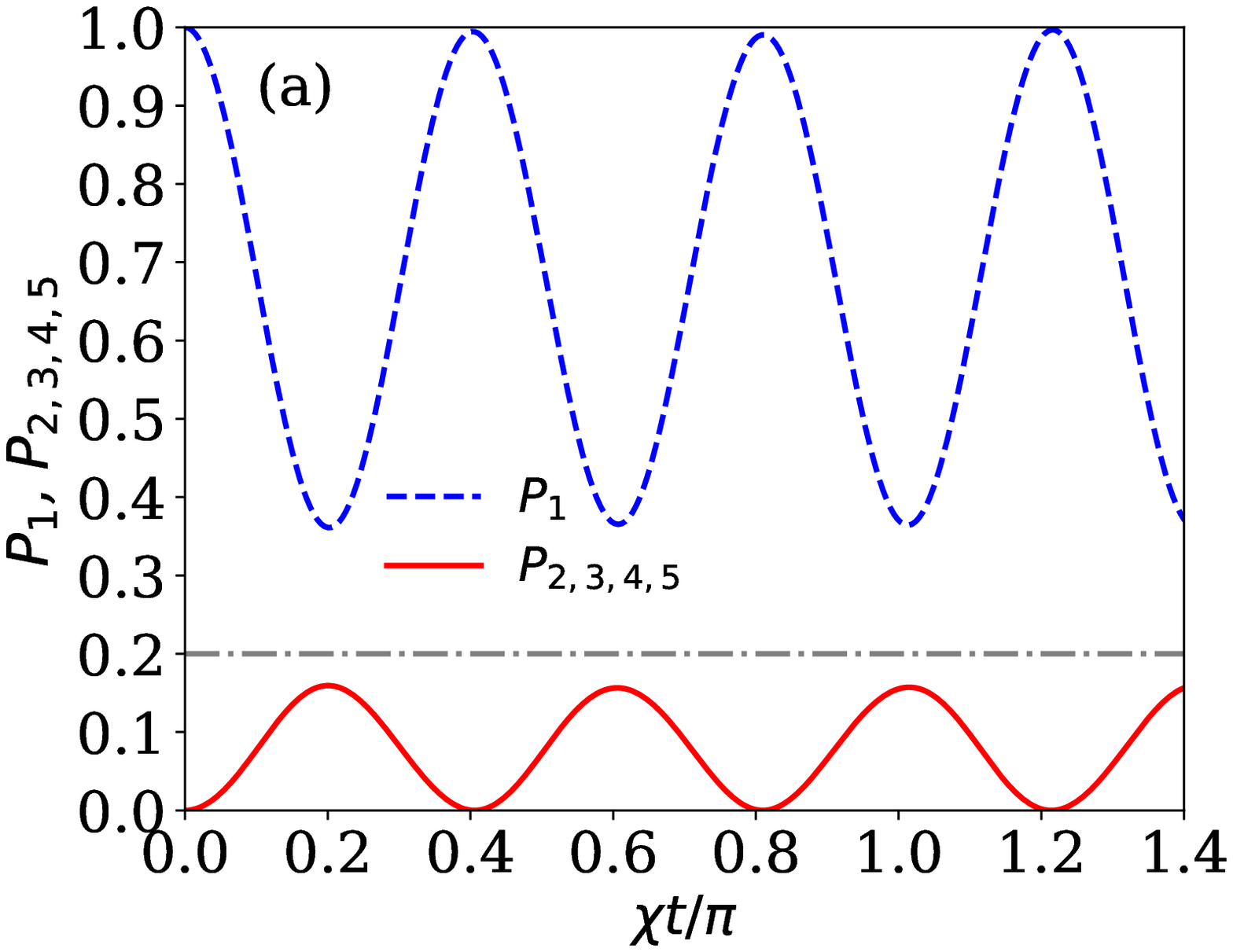}\\
\includegraphics[width=3.25 in]{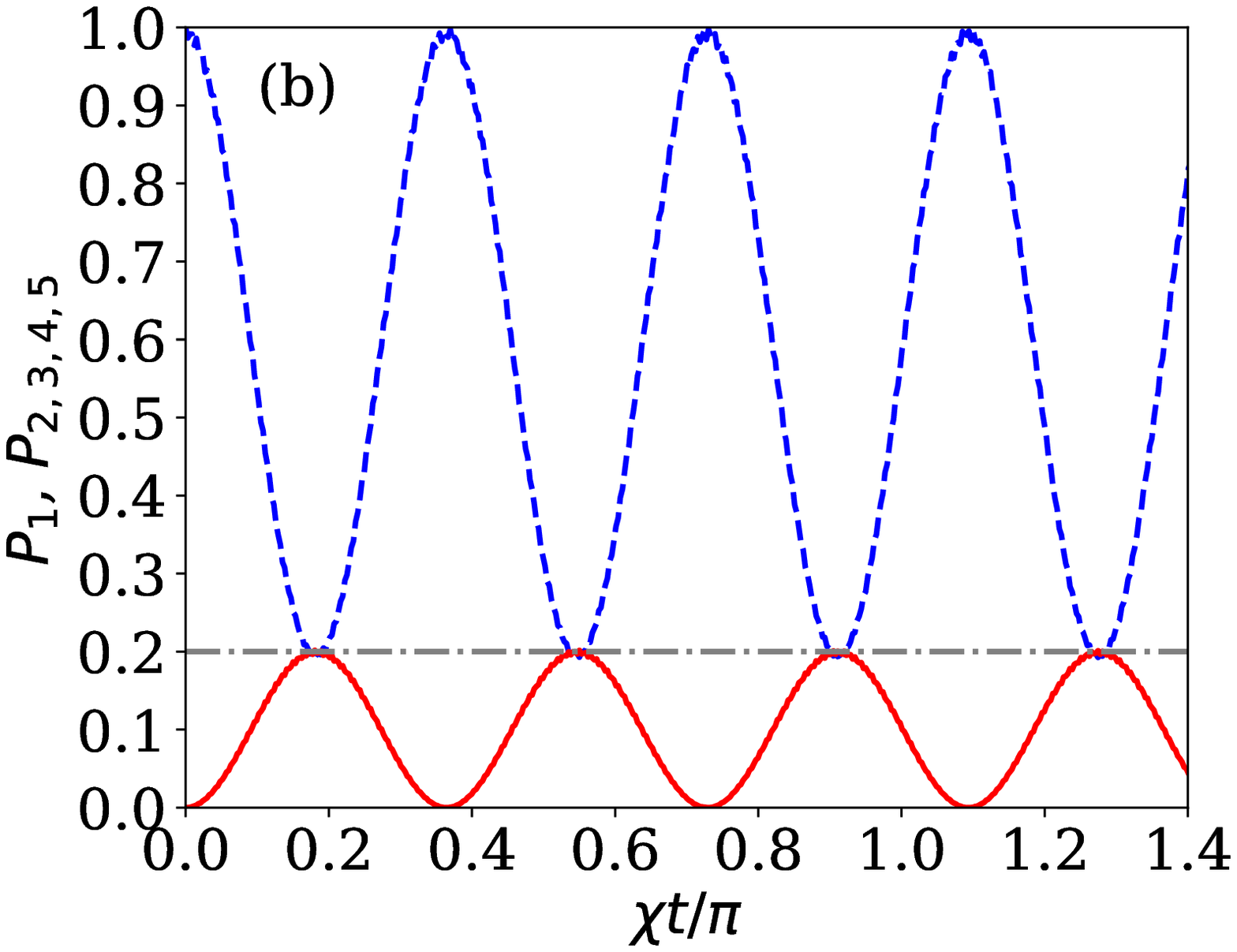} 
\end{tabular}
\caption{(a) Illustration for the time evolution of the photon population in six-resonator subsystem when system is assumed to be in the initial state $\left\vert 1_{1} 0_{2} 0_{3} 0_{4} 0_{5} \right\rangle \left\vert 0_{0} \right\rangle$.
A gap appears in the photon state population $P_{1}$ (dashed-blue curve) and $P_{2,3,4,5}$ (red-solid curve).
(b) Dynamics of the photon population when coupling strength of resonator $R_{1}$ changed to a higher optimum value of $g_{1}=2 \pi \times 62.5$~MHz while rest of the coupling strengths are unchanged i.e., $g=g_{2}=...=g_{5}=2 \pi \times 50$~MHz.
As a result, population gape gets filled and the dynamics of the system can nearly produce five-qubit W-state on photon states as marked by the horizontal dashed-dotted line.}
\label{fig:pop5}
\end{figure}

Next, we extend the scheme for time dynamics and generation of entanglement to an arbitrary $n$-distant resonators.
We consider $n$ resonators $R_{j}$ ($j=1,2,3,...,n$) coupled to a common resonator $R_{0}$ in the dispersive regime having identical coupling constant. 
In this case, following the similar unitary transformation (See Eq.~(\ref{U1})) up to $j=n$, the effective Hamiltonian for $n+1$-resonator subsystem can be written in the following form:
\begin{equation}
 H_{\rm eff}^{n} = \sum_{i \neq j, i<j}^{n} \chi_{ij} \left(b_{i} b_{j}^{\dag} e^{i \delta_{ij} t}+H.c\right).
\label{hn}
\end{equation}
Here, we assume that the initial state of $n+1$-resonator subsystem is $\left\vert 1_{1} 0_{2} 0_{3} ... 0_{n} \right\rangle \left\vert 0_{0} \right\rangle$ and all the distant resonators are resonant to each other i.e., equally detuned from the common resonator $R_{0}$.
Under this condition, the state of $n+1$-resonator subsystem evolves at time $t$ with coefficients
\begin{subequations}
\label{ampn}
\begin{align}
C_{1} (t)& =\frac{1}{n} \left[(n-1) \cos \chi t + \cos (n-1) \chi t\right] \nonumber\\
				 & + \frac{i}{n} \left[(n-1) \sin \chi t - \sin (n-1) \chi t \right],\\
C_{2} (t)& =C_{3}=...=C_{n}(t) =\frac{1}{n} \left[- \cos \chi t + \cos (n-1) \chi t\right]\nonumber\\
         & - \frac{i}{n} \left[\sin \chi t + \sin (n-1) \chi t \right],
\end{align}
\end{subequations}
which reduce to Eq.~(\ref{amp}) i.e., four-resonator subsystem for $n=3$. 
As an example, we consider the case of five-resonator subsystem composed of resonators $R_{1}$, $R_{2}$, $R_{3}$, and $R_{4}$ coupled to $R_{0}$. 
The initial state of this system is assumed to be $\left\vert 1_{1} 0_{2} 0_{3} 0_{4} \right\rangle \left\vert 0_{0} \right\rangle$.
In Fig.~\ref{fig:pop4}, dashed-blue curve shows the time evolution of the population $P_{1}$ of state $\left\vert 1_{1} 0_{2} 0_{3} 0_{4} \right\rangle \left\vert 0_{0} \right\rangle$ while solid-red curve denotes the time evolution of the population $P_{2,3,4}$ of state $\left\vert 0_{1} 1_{2} 0_{3} 0_{4} \right\rangle \left\vert 0_{0} \right\rangle$, $\left\vert 0_{1} 0_{2} 1_{3} 0_{4} \right\rangle \left\vert 0_{0} \right\rangle$, and $\left\vert 0_{1} 0_{2} 0_{3} 1_{4} \right\rangle \left\vert 0_{0} \right\rangle$. 
The parameters taken here are similar to the case considered in Fig.~\ref{fig:pop3}. 
The time evolution of the photon state clearly produces four-qubit W-state at interaction times $\chi t / \pi= 1/4,3/4, 5/4$. By setting the operation time $\chi t= \pi /4$, one can get the following four-qubit entangled state: 
\begin{align}
\left\vert \psi_{4} \right\rangle_{\pi/4} & = \frac{1+i}{2 \sqrt{2}} ( \left\vert 1_{1} 0_{2} 0_{3} 0_{4} \right\rangle - \left\vert 0_{1} 1_{2} 0_{3} 0_{4} \right\rangle  \nonumber\\
																					& - \left\vert 0_{1} 0_{2} 1_{3} 0_{4} \right\rangle- \left\vert 0_{1} 0_{2} 0_{3} 1_{4} \right\rangle),
\label{w4}
\end{align}
which is similar to the prototype four-qubit W-sate.

Next, we consider the case of six-resonator subsystem composed of resonators $R_{1}$, $R_{2}$, $R_{3}$, $R_{4}$, and $R_{5}$ coupled to $R_{0}$ with initial state $\left\vert 1_{1} 0_{2} 0_{3} 0_{4} 0_{5} \right\rangle \left\vert 0_{0} \right\rangle$.
In Fig.~\ref{fig:pop5}(a), we plot time evolution of the population $P_{1}$ of state $\left\vert 1_{1} 0_{2} 0_{3} 0_{4} 0_{5} \right\rangle \left\vert 0_{0} \right\rangle$ (dashed-blue curve) and $P_{2,3,4,5}$ of states $\left\vert 0_{1} 1_{2} 0_{3} 0_{4} 0_{5} \right\rangle \left\vert 0_{0} \right\rangle$, $\left\vert 0_{1} 0_{2} 1_{3} 0_{4} 0_{5} \right\rangle \left\vert 0_{0} \right\rangle$, $\left\vert 0_{1} 0_{2} 0_{3} 1_{4} 0_{5} \right\rangle \left\vert 0_{0} \right\rangle$, and $\left\vert 0_{1} 0_{2} 0_{3} 0_{4} 1_{5} \right\rangle \left\vert 0_{0} \right\rangle$ (denoted by solid-red curves). 
The period of oscillation for the photon state population gets shorter as compared to the four and five-resonator-qubit subsystem. 
Furthermore, the peak amplitude of the oscillation for the photon states also decreases as the number of resonators increase. As a result, a gap appears in the photon state population $P_{1}$ and $P_{2,3,4,5}$, which is clearly visible in Fig.~\ref{fig:pop5}(a). 
In order to obtain the prototype five-qubit W-state with equal probability amplitudes of $1/5$, one needs crossing point of the photon state population. For this purpose, we numerically solve the density matrix equation $d \rho / dt = -i \left[H, \rho \right]$ by considering full Hamiltonian given by Eq.~(\ref{H}) with $g_{1}$ different than $g=g_{2}=,..,=g_{5}$.
The results of our simulations are shown in Fig.~\ref{fig:pop5}(b) with $g_{1}=2 \pi \times 62.5$~MHz while rest of the parameters are the same as considered in Fig.~\ref{fig:pop3}. 
Figure.~\ref{fig:pop5}(b) clearly reveals that the dynamics of the system can nearly evolve to a five-qubit W-state at four different interaction times $\chi t / \pi$ as indicated by the dashed-dotted horizontal line.

\begin{figure}[tbp]
\begin{tabular}{@{}cccc@{}}
\includegraphics[width=3.25 in]{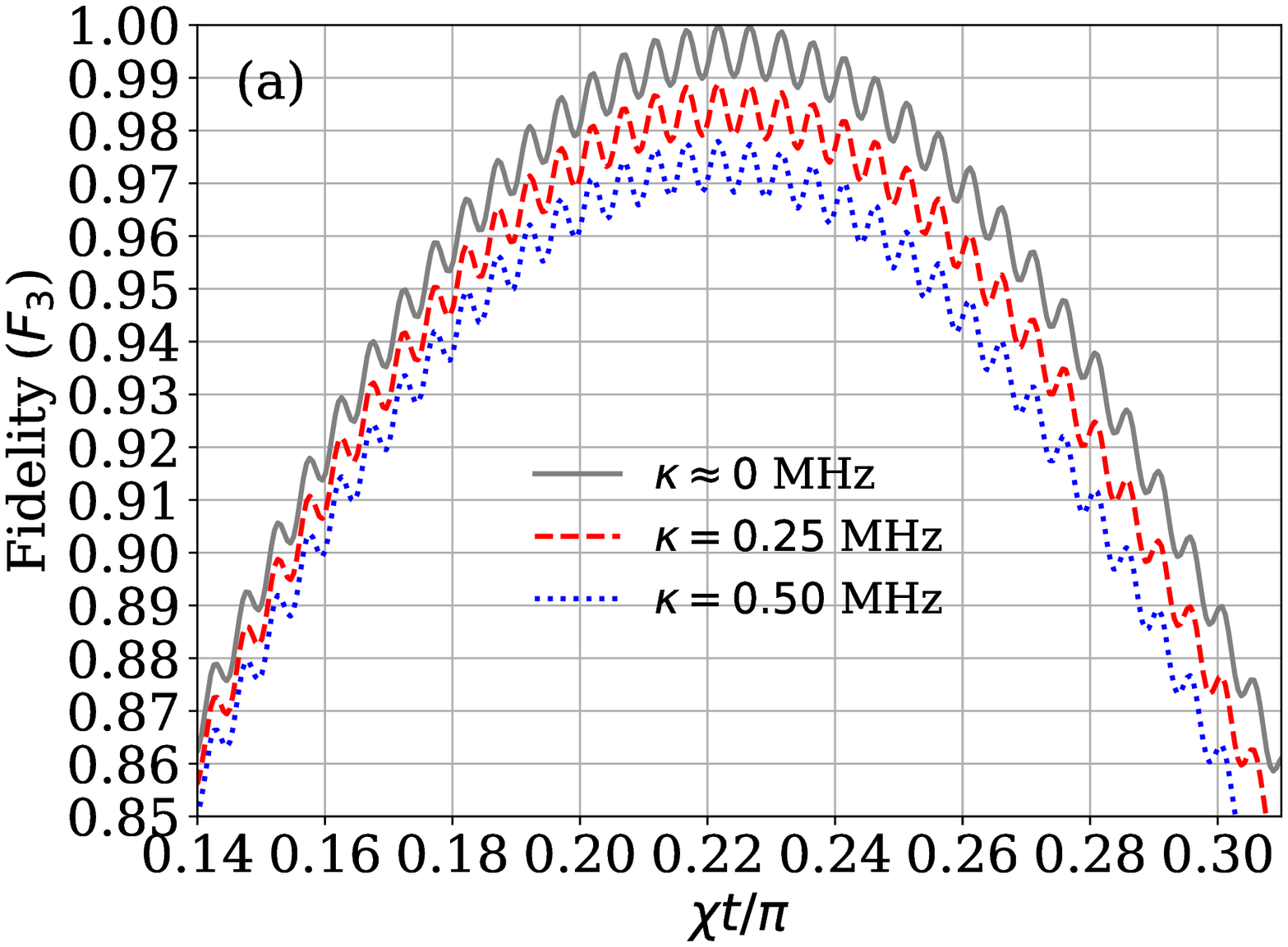}\\
\includegraphics[width=3.25 in]{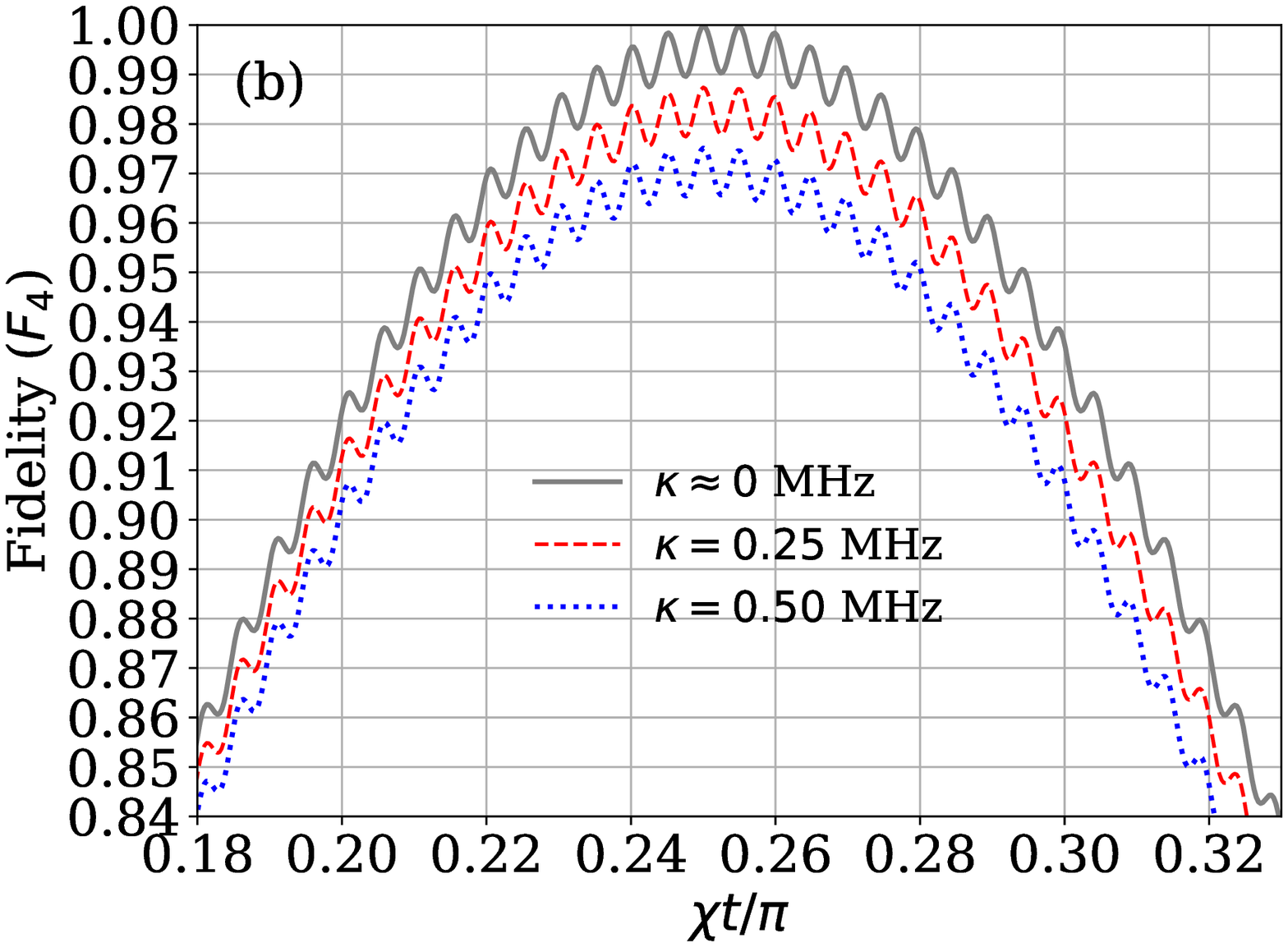} 
\end{tabular}
\caption{(a) Numerical illustration for the fidelity of the three-qubit W-state in the four-resonator subsystem as a function of operation time under photon decay rates of resonators.
Here, solid black curve, dashed-red curve, and dotted-blue curve corresponds to $\kappa=0$~MHz, $\kappa=0.25$~MHz (life time $\kappa^{-1}=4~\mu$s), and $\kappa=0.5$~MHz (life time $\kappa^{-1}=2~\mu$s), respectively. 
The rest of the parameters are the same as used in Fig~\ref{fig:pop3}.
(b) Fidelity of the four-qubit W-state in the five-resonator subsystem as a function of operation time for the same photon decay rates as used in Fig.~\ref{fig:fid3}(a).
}
\label{fig:fid3}
\end{figure}

\section{Possible experimental implementation}
In general, our scheme for generating W-state is possible using different kind of quantum system that follows the ladder algebra of quantum harmonic oscillator i.e., $[a, a^{\dag}]=1$, and dispersive interaction regime. Such as photon states inside the optical cavities or superconducting resonators~\cite{ming45,maxwell}.
However, considering two-level neutral or artificial atoms (Trasmon or Cooper Pair Box etc) as a qubit coupled to common resonator~\cite{helmer,gao,wei} may not work. Because, in case of a two-level atom, ladder algebra follows $[\sigma^{-}, \sigma^{+}] \neq 1$, which may not result in the adiabatic elimination of the common resonators with dispersive interaction.

Recently, quantum information processing using microwave photons in a one-dimensional superconducting resonator has been focused due to high-quality factor and large zero-point electric fields.
Some interesting outcomes have been presented such as catching and releasing of microwave photon states~\cite{ming45} and quantum superposition of a single microwave photon~\cite{ming46}. 
Therefore, we consider the experimental feasibility of our results on microwave photon states in a system of superconducting resonators as shown in Fig~\ref{fig:sys}.
\begin{figure}[tbp]
\includegraphics[width=3.25 in]{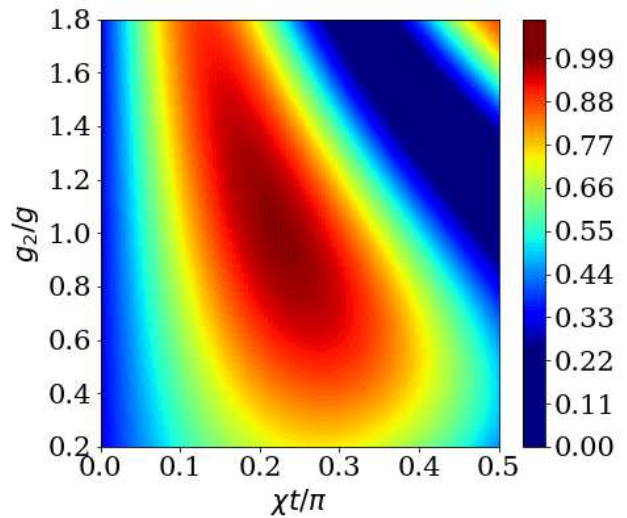}
\caption{Fidelity of three-qubit W-state versus $g_{2}/g$ and operation time $\chi t / \pi$ for $\kappa=0.10$~MHz in four-resonator subsystem. The rest of the parameters are the same as used in Fig~\ref{fig:fid3}. The small dark shaded contour corresponds to maximum fidelity.}
\label{fig:fid3d}
\end{figure}

Imperfections in the implementation can affect the time dynamics and hence the realization of the W-state entanglements. 
The possible sources of imperfections could be the finite lifetime of the photons in the superconducting resonators, inhomogeneous coupling strength $g_{j}$, direct resonator-resonator interactions due to mutual inductance or capacitance, and noisy initial state preparation.
We first consider the effects of photon lifetime $\kappa$ in the four-resonator subsystem.
The time evolution of the four-resonator subsystem under the influence of photon decay rates can be described by the master equation~\cite{lambro}
\begin{equation}
 \frac{d \tilde{\rho}}{d t} = -i \left[H, \tilde{\rho} \right]+ \kappa_{0}L\left[ a \right] + \kappa_{j}L\left[ b_{j} \right]
\label{master}
\end{equation}
where, $j=1,2,3$, $\kappa_{0}$ is decay rate of superconducting resonator bus and $\kappa_{j}$ is the decay rate of the $j^{\rm th}$ superconducting resonator.
The Liouvillian of an operator $\xi \in (a,b_{j})$ is $L\left[ \xi \right]= \xi \rho \xi^{\dag}- \xi^{\dag} \xi \rho /2 + \rho \xi^{\dag} \xi/2$, which describes the photon decay through the resonator.
In Eq.~(\ref{master}), $\tilde{\rho} (t)$ is the final density operator of the system when the operation is performed under photon leakage from the superconducting resonator.
We numerically obtained the density operator $\tilde{\rho}(t)$~\cite{qutip} with initial state $\left\vert \psi_{1} \right\rangle= \left\vert 1_{1} 0_{2} 0_{3} \right\rangle \left\vert 0_{0} \right\rangle$ and for Hamiltonian $H$ given by Eq.~(\ref{H}).
We first numerically simulate the population of microwave photons $P_{1}$, and $P_{2,3}$ as a function of $\chi t / \pi$ as shown in Fig.~\ref{fig:pop3} by dotted curves.
Here, we adopted $\kappa=\kappa_{0}=\kappa_{1}=\kappa_{2}=\kappa_{3}=0.5$~MHz (lifetime of 2~$\mu$s) while the rest of the parameters are the same as used in Fig.~\ref{fig:pop3}.
The damping occurs in the oscillation of $P_{1}$ and $P_{2,3}$ due to the photon decay through the resonator. 
Its effect becomes more prominent at longer operation times as clearly visible by dotted curves in Fig.~\ref{fig:pop3}. 
Therefore, choosing shorter operation time (first crossing point) is more feasible to obtain a W-state.

Next, we present the results of our numerical simulations for the fidelity $F_{3}$ of the entangled state $\left\vert \psi_{3} \right\rangle_{0.22}$ generated from the initial state $\left\vert \psi_{1} \right\rangle$ as a function of operation time using the definition
\begin{equation}
 F_{3} = \left\langle \psi_{3} \left| \tilde{\rho}(t) \right| \psi_{3} \right\rangle.
\label{fid3}
\end{equation}
Here, $\left| \psi_{3} \right\rangle$ is the output state of an ideal system and $\tilde{\rho} (t)$ is the final density operator under photon leakage from the superconducting resonator.
Figure~\ref{fig:fid3}(a) shows the result of simulation obtained from Eq.~(\ref{fid3}) for different photon decay rates~\cite{cp2018}.
Here, solid black, dashed-red, and dotted-blue curves correspond to $\kappa=0$~MHz, $\kappa=0.25$~MHz (life time $\kappa^{-1}=4~\mu$s), and $\kappa=0.5$~MHz (life time $\kappa^{-1}=2~\mu$s), respectively.
The rest of the parameters are the same as used in Fig.~\ref{fig:pop3}.
Figure~\ref{fig:fid3}(a) clearly shows that the fidelity varies with operation time for different decay rates of the resonators. 
Here, fidelity is maximum for $\kappa=0$~MHz (ideal case), 98.7$\%$ with $\kappa=0.25$~MHz, and 97.7$\%$ with $\kappa=0.5$~MHz for optimum operation time $\chi t / \pi \approx 0.22$.

Next, we numerically simulate the fidelity $F_{4}$ of the W-state generated in five-resonator subsystem using the expression
\begin{equation}
 F_{4} = \left\langle \psi_{4} \left| \tilde{\rho} (t) \right| \psi_{4} \right\rangle.
\label{fid4}
\end{equation}
Here, $\left| \psi_{4} \right\rangle$ is the W-state given by Eq.~(\ref{w4}) under ideal condition and $\tilde{\rho} (t)$ is the final density operator under photon leakage from the superconducting resonator.
The results of our numerical simulations are shown in Fig.~\ref{fig:fid3}(b). 
Again fidelity $F_{4}$ varies with time at different decay rates and the maximum value of the fidelity is obtained around operation time $\chi t / \pi \approx 1/4$. 
Here, maximum fidelity can approximately reach 100$\%$ with $\kappa=0$~MHz (ideal case), 98.4$\%$ with $\kappa=0.25$~MHz, and 97.4$\%$ with $\kappa=0.5$~MHz.
It is evident from the results of Figs.~\ref{fig:fid3}(a) and (b) that W-state can be generated in a single step with high fidelity.
The small oscillations in Figs.~\ref{fig:fid3}(a) and (b) arises due the interactions between $R_{j}$ and $R_{0}$. These oscillations can be further minimized by increasing the detuning $\Delta$ or reducing the coupling strengths $g$.

Furthermore, so far, we have assumed identical coupling constant $g_{j}$ between each distant resonator $R_{j}$ and common superconducting resonator $R_{0}$ in our analysis. 
Needless to say, this assumption would not be perfectly met in an experimental setup.
The coupling inaccuracies can induce small difference in the value of $g_{j}$ for each resonator. 
In order to get further insight, next, we consider the fidelity of W-state against the inhomogeneous coupling strength.
For simplicity, we consider four-resonator subsystem in which we vary $g_{2}$ relative to the other couplings i.e., $g=g_{1}=g_{3}$.
In Fig.~\ref{fig:fid3d}, we show a contour plot of fidelity for three-qubit W-state as function of $\chi t / \pi$ and $g_{2} / g$ for photon decay rate of $\kappa=0.10$~MHz (life time $\kappa^{-1}=10~\mu$s).
The small dark-shaded region of the contour in Fig.~\ref{fig:fid3d} shows that the fidelity can reach up to approximately 99$\%$ when $g_{2} \sim g$ for operation time around $\chi t / \pi \approx 0.22$.
\begin{figure}[tbp]
\includegraphics[width=3.25 in]{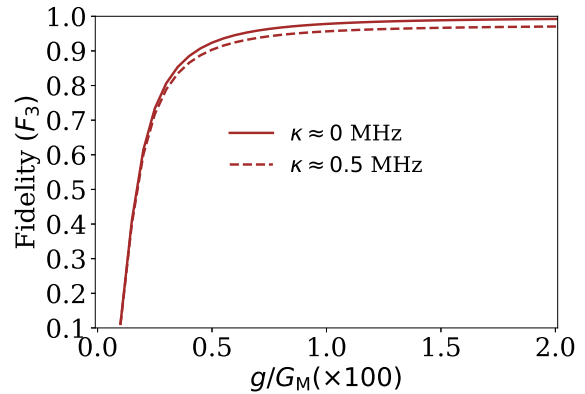}
\caption{Fidelity of W-state versus coupling ratio $g/G_{\rm M}$ for $\kappa=0$~MHz (solid curve) and $\kappa=0.5$~MHz (dashed curve) in four-resonator subsystem.}
\label{fig:fidmutual}
\end{figure}
\begin{figure}[tbp]
\includegraphics[width=3.25 in]{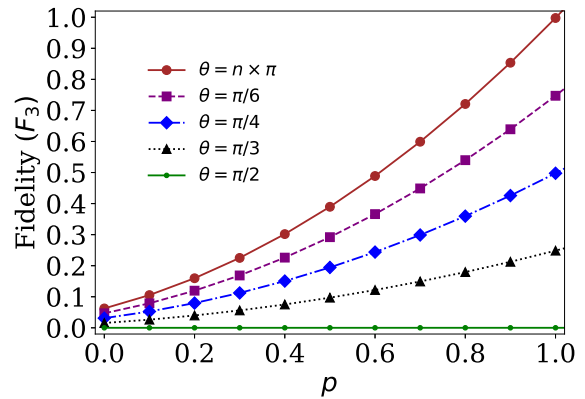}
\caption{Fidelity of W-state versus noise parameter $p$ for the initial Werner's mixed state at different overlap angles in four-resonator subsystem. Fidelity approaches to unity when $p=1$ and $\theta = n \times \pi$ (such that $n=0,1,2,...$).}
\label{fig:fidmix}
\end{figure}

In our scheme, $R_{j}$ resonators are coupled through resonator-bus-mediated mechanism instead of direct resonator-resonator coupling. This mediated mechanism may allows neglecting any parasitic mutual inductance or capacitance.
Further more, the spacing between each superconducting resonators $R_{j}$ can be kept large enough to avoid interaction among them due to mutual capacitance and mutual inductance.
However, in a realistic situation, a direct resonator-resonator coupling my still appear. In order to study the effects of this direct interaction in the generation of W-state, we added the additional interaction term~\cite{fazal,fazal24}
\begin{equation}
H_{I,M}=G_{\rm M} (b_{1} b_{2}^{\dag} + b_{2} b_{3}^{\dag} + H.c.,)
\label{Hm}
\end{equation}
in Eq.~(\ref{int}) for four-resonator subsystem. 
Here, $G_{\rm M}$ is the direct resonator-resonator coupling parameter, which includes the effects of mutual inductance or capacitance.
Equation~(\ref{Hm}) shows that the first resonator gets decoupled from the third resonator if we only consider the direct resonator-resonator interaction ($G_{\rm M} \neq 0$) instead of resonator-bus-mediated mechanism i.e., $g_{j}=0$ in Eq.~(\ref{int}). As a result time dynamics will change significantly.
Now applying the unitary transformation Eq.~(\ref{U1}) breaks down the Schrieffer-Wolff Transformation~\cite{swtrans} (See Appendix~\ref{A}). As a result, calculations become extremely tedious to obtain an effective Hamiltonian with renormalization of parameters $\chi$'s in simplified form.
Therefore, we numerically obtained the fidelity $F_{3}$ as a function of coupling ratio $g/G_{\rm M}$ as shown in Fig.~(\ref{fig:fidmutual}). 
The results in Fig.~(\ref{fig:fidmutual}) clearly show that high fidelity can be obtained when direct resonator-resonator coupling $G_{\rm M}$ is approximately $100$ time weaker as compared with the resonator-bus coupling $g$.

Furthermore, initial state $\left\vert \psi_{1} \right\rangle$ can be realized with high fidelity by catching the photons from the transmission lines by turning on the coupling capacitance~\cite{ming45}. 
However, real situation is that the initial state will be a complicated mixed state. 
Next we are interested to study how the fidelity of the W-state preparation is affected if one includes noise in the initial state preparation.
For simplicity, we consider the initial state which is similar to noisy Werner’s mixed State ~\cite{werner,werner96,werner98} in four resonator subsystem given by
\begin{equation}
 \rho_{i} = p \left\vert \Phi_{i}  \right\rangle  \left\langle \Phi_{i} \right|-\frac{1-p}{4} I_{8}.
\label{wernstate}
\end{equation}
Here, $0 \leq p \leq 1$ is a noise parameter which includes imperfection due to time controlled capture and release of microwave photons and unwanted hoping of microwave photons among resonators.
An arbitrary pure state is given by $\left\vert \Phi_{i}  \right\rangle = \left(\cos \theta \left\vert 1_{1} 0_{2} \right\rangle +i \sin \theta  \left\vert 0_{1} 1_{2} \right\rangle\right) \left\vert 0_{3} 0_{0} \right\rangle$ and $I_{8}$ is an identity operator.
The parameters $p$ and $\theta$ measure the overlap of the mixed state Eq.~(\ref{wernstate}) with an ideal initial pure state $\left\vert \psi_{1}  \right\rangle $.
In Fig.~\ref{fig:fidmix}, we show the fidelity $F_{3}$ as a function of noise parameter $P$ for different values of angles $\theta$. 
As expected, when $p=1$ and $\theta = n \times \pi$ (such that $n=0,1,2,...$), the initial state becomes $\rho_{1} = \left\vert \psi_{1}  \right\rangle  \left\langle \psi_{1} \right|$ and fidelity approaches to unity.

The influence of superconducting ring couplers may also arise in time dynamics and entanglement. However, it can be neglected because their frequencies are designed at large detuning with those of superconducting resonators~\cite{ming57}.
Finally, the quality factor $Q$ for a resonator describes the life time of photons inside the resonator. 
Higher $Q$ value indicate a lower rate of photon loss from the resonator.
It may be worth mentioning that the quality factor for a one-dimensional superconducting resonator which has been achieved in some recent experiment is of the order of $Q\sim 2 \times 10^{6}$~\cite{ming59,cp16}. 
Using this quality factor, the estimated lifetime of the photons in a superconducting resonator is $\kappa^{-1} = 1/(\omega_{r}/Q) \approx 55~\mu$s, which is sufficiently longer than the operation time in our scheme. 
The analysis here implies that the high fidelity implementation of multiqubit W-State is feasible in distant resonators in a single-step.

\section{Conclusion}
In conclusion, we have studied the time dynamics of the photon state population in distant resonators. 
Each resonators $R_{j}$ is coupled to a common quantum bus resonator $R_{0}$.
In our scheme, all resonator $R_{j}$ have identical frequency detuning from a common bus resonator $R_{0}$.
In order to satisfy the dispersive regime, the frequency detuning is kept larger than the coupling strengths to a common bus resonator.
In the dispersive regime, the time dynamics of three and four resonators coupled to a common bus evolve to three and four-qubit W-state, respectively. 
A population gap appears when five or higher number of resonators are coupled to a common bus resonator. 
We show that the dynamics of the system can still produce multiqubit W-state by increasing the coupling strength of the first resonator to a larger optimum value as compared to the rest of the coupling strengths.
Our numerical simulations in a system of superconducting microwave resonators show the high fidelity generation of single-step W-state against the possible sources of imperfections in a realistic situation.
Since it is possible to achieve high Q-value resonators, catching and releasing a photon in resonator from a transmission line in the experiment, therefore, our results may stimulate experimental activities in the near future.

\begin{appendices}
\numberwithin{equation}{section}
\section{Derivation of Dispersive Hamiltonian}
\label{A}
Here we list some useful commutator relations, which we have used to diagonalize the Hamiltonian given by the Eq.~(\ref{Hs})
\begin{subequations}
\label{comlist}
\begin{align}
\left[a b_{j^{\prime}}^{\dag} - a^{\dag}b_{j^{\prime}}, a^{\dag} a \right] &=-\left(a^{\dag}b_{j} + a b_{j}^{\dag} \right),\\
\left[a b_{j^{\prime}}^{\dag} - a^{\dag}b_{j^{\prime}}, b_{j}^{\dag} b_{j} \right] & =\left(a^{\dag}b_{j} + a b_{j}^{\dag}  \right)  \delta_{j j^{\prime}},\\
\left[a b_{j^{\prime}}^{\dag} - a^{\dag}b_{j^{\prime}}, a^{\dag} b_{j} + a b_{j}^{\dag} \right] & =\left(b_{j^{\prime}}^{\dag} b_{j} + b_{j^{\prime}} b_{j}^{\dag} \right) \nonumber\\
				 & - \left( a^{\dag} a + a a^{\dag} \right) \delta_{j j^{\prime}}.
\end{align}
\end{subequations}

In order to diagonalize the Hamiltonian in Eq.~(\ref{Hs}), we consider the unitary transformation of Eq.~(\ref{U1}) as $U=e^{S}$, with 
\begin{equation}
S=\sum_{j^{\prime}=1}^{3} \frac{g_{j^{\prime}}}{\Delta_{j^{\prime}}} \left(a b^{\dag}_{j^{\prime}}-a^{\dag} b_{j^{\prime}} \right).
\label{sw}
\end{equation}
By considering $H_{s}=H_{0,s}+H_{I,s}$, the unitary transformation $U=U H_{S} U^{\dag}$ yields
\begin{align}
H^{\prime} & = H_{0,s}+H_{I,s}+[S,H_{0,s}]+[S,H_{I,s}]\nonumber\\
					 & + \frac{1}{2} [S,[S,H_{0,s}]]+\frac{1}{2} [S,[S,H_{I,s}]]+...,
\label{BCH0}
\end{align}
where,
\begin{align}
\left[S, H_{0,s}\right]& =\sum_{j^{\prime}=1}^{3} \frac{g_{j^{\prime}} \Omega_{0}}{\Delta_{j^{\prime}}}  \left[ \left(a b^{\dag}_{j^{\prime}}-a^{\dag} b_{j^{\prime}} \right), a^{\dag}a \right]\nonumber\\ 
                       & + \sum_{j j^{\prime}=1}^{3} \frac{g_{j^{\prime}} \omega_{j}}{\Delta_{j^{\prime}}} \left[  \left(a b^{\dag}_{j^{\prime}}-a^{\dag} b_{j^{\prime}} \right),  b^{\dag}_{j} b_{j} \right].
\end{align}
On using Eq.~(\ref{comlist}), we obtain
\begin{align}
\left[S, H_{0,s}\right]& =\sum_{j^{\prime}=1}^{3} \frac{g_{j^{\prime}} \Omega_{0}}{\Delta_{j^{\prime}}}  \left[ - \left(a^{\dag} b_{j^{\prime}} +a b^{\dag}_{j^{\prime}} \right) \right]\nonumber\\
                       & + \sum_{j j^{\prime}=1}^{3} \frac{g_{j^{\prime}} \omega_{j}}{\Delta_{j^{\prime}}} \left[ \left(a^{\dag} b_{j^{\prime}} +a b^{\dag}_{j^{\prime}} \right) \delta_{j j^{\prime}}\right], \nonumber\\
											& = - \sum_{j=1}^{3} \frac{g_{j} }{\Delta_{j}} \left(a^{\dag} b_{j} +a b^{\dag}_{j} \right) \times \left( \Omega_0 - \omega_{j} \right).
\end{align}
Since, $\Delta_{j}=\Omega_0 - \omega_{j}$, therefore, we have
\begin{equation}
\left[S, H_{0,s}\right] = -H_{I,s}.
\label{coms}
\end{equation}
Now using Eq.~(\ref{coms}) in Baker–Campbell–Hausdorff (BCH) formula given by Eq.~(\ref{BCH0}), we obtain
\begin{equation}
H^{\prime} =H_{0,s}+\frac{1}{2} [S,H_{I,s}]+...,
\label{BCH1}
\end{equation}
which is the Schrieffer-Wolff Transformation~\cite{swtrans}. One only needs to calculate the commutator

\begin{align}
\left[S, H_{I,s}\right]& = \sum_{j j^{\prime}=1}^{3} \frac{g_{j^{\prime}} g_{j}}{\Delta_{j^{\prime}}} \left[ \left(  a b^{\dag}_{j^{\prime}}-a^{\dag} b_{j^{\prime}} \right), \left(a^{\dag} b_{j} +a b^{\dag}_{j} \right) \right].
\end{align}
By using the commutator relations from Eq.~(\ref{comlist}), we obtain
\begin{align}
\left[S, H_{I,s}\right] & =\sum_{j j^{\prime}=1}^{3} \frac{g_{j^{\prime}} g_{j}}{\Delta_{j^{\prime}}} \left[ \left(  b^{\dag}_{j^{\prime}} b_{j}+b_{j^{\prime}} b_{j}^{\dag} \right)- \left( a^{\dag} a + a a^{\dag} \right) \delta_{j j_{\prime}} \right]\nonumber\\
												& = \sum_{j=1}^{3} \frac{2 g_{j}^{2}}{\Delta_{j}} \left( b_{j}^{\dag} b_{j} - a^{\dag} a \right)\nonumber\\
												& + \sum_{j \neq j^{\prime}=1}^{3} \frac{g_{j^{\prime}} g_{j}}{\Delta_{j^{\prime}}} \left( b^{\dag}_{j^{\prime}} b_{j}+b_{j^{\prime}} b_{j}^{\dag} \right).
\label{swcom}
\end{align}
By using Eq.~(\ref{swcom}) in Eq.~(\ref{BCH1}), we obtain 
\begin{align}
H^{\prime} & = a^{\dag} a \left(\Omega_0 - \sum_{j=1}^{3} \frac{g_{j}^{2}}{\Delta_{j}} \right) + \sum_{j=1}^{3} b_{j}^{\dag} b_{j} \left(\omega_{j} + \frac{g_{j}^{2}}{\Delta_{j}} \right)\nonumber\\
					 & + \sum_{j \neq j^{\prime}=1}^{3} \frac{g_{j^{\prime}} g_{j}}{2 \Delta_{j^{\prime}}} \left( b^{\dag}_{j^{\prime}} b_{j}+b_{j^{\prime}} b_{j}^{\dag} \right),
\end{align} 
which is diagonal upto $g_{j}^{2} / \Delta_{j}$ and in agreement with Eq.~(\ref{hi}).


\end{appendices}



\end{document}